\documentclass{jpsj-suppl}
\usepackage{txfonts} 

\def\cH{{\mathcal H}}
\def\cHeff{{\mathcal H}_{\rm eff}}
\def\rd{{\rm d}}
\def\rr{{\rm r}}
\def\rS{{\rm S}}
\def\ket#1{|#1\rangle}
\def\up{\uparrow}
\def\dn{\downarrow}
\def\Up{\Uparrow}
\def\Dn{\Downarrow}
\def\vS{{\mib S}}
\def\vT{{\mib T}}
\title{Inversion Phenomena of the Anisotropies of the Hamiltonian and the Wave Function in
Quantum Spin Chains
}

\author{Kiyomi \textsc{Okamoto}}

\inst{Department of Physics, Tokyo Institute of Technology, Ookayama, Meguro-ku, Tokyo 152-8551, Japan}

\email{kokamoto@phys.titech.ac.jp}

\recdate{July 15, 2013}

\abst{
We investigate the inversion phenomena between the $XXZ$ anisotropies
of the Hamiltonian and the wave function in quantum spin chains.
We focus on the $S=1/2$ geometrically frustrated 3-leg ladder system with the $XXZ$
interaction anisotropy.
By use of the degenerate perturbation theory from the strong rung coupling limit,
we have obtained the ground-state phase diagram. 
In some parameter regions,
the Tomonaga-Luttinger spin liquid state is realized in spite of the Ising-like anisotropy,
and the N\'eel state in spite of the $XY$-like anisotropy.
}

\kword{quantum spin, low dimension, spin ladder, quantum phase transition}

\begin{document}
\maketitle

\section{Introduction}

In the ground state problem of the antiferromagnetic quantum spin chains
with the $XXZ$ interaction anisotropy
$\Delta \equiv J_z/J_\perp$,
the Ising-like anisotropy $\Delta>1$ is usually favorable to the N\'eel state
and the $XY$-like anisotropy $\Delta<1$ to the Tomonaga-Luttinger spin liquid (TLL) state.
For instance, the ground state of the simple $S=1/2$ antiferromagnetic $XXZ$ chain is either of the N\'eel
state or
the TLL state according as $\Delta > 1$ or $\Delta \le 1$.
A similar situation is found in the simple $S=1$ $XXZ$ chain,
although the Haldane phase exists between the N\'eel phase and the TLL phase.

\begin{figure}[ht]
       \centerline{
          \scalebox{0.45}{\includegraphics{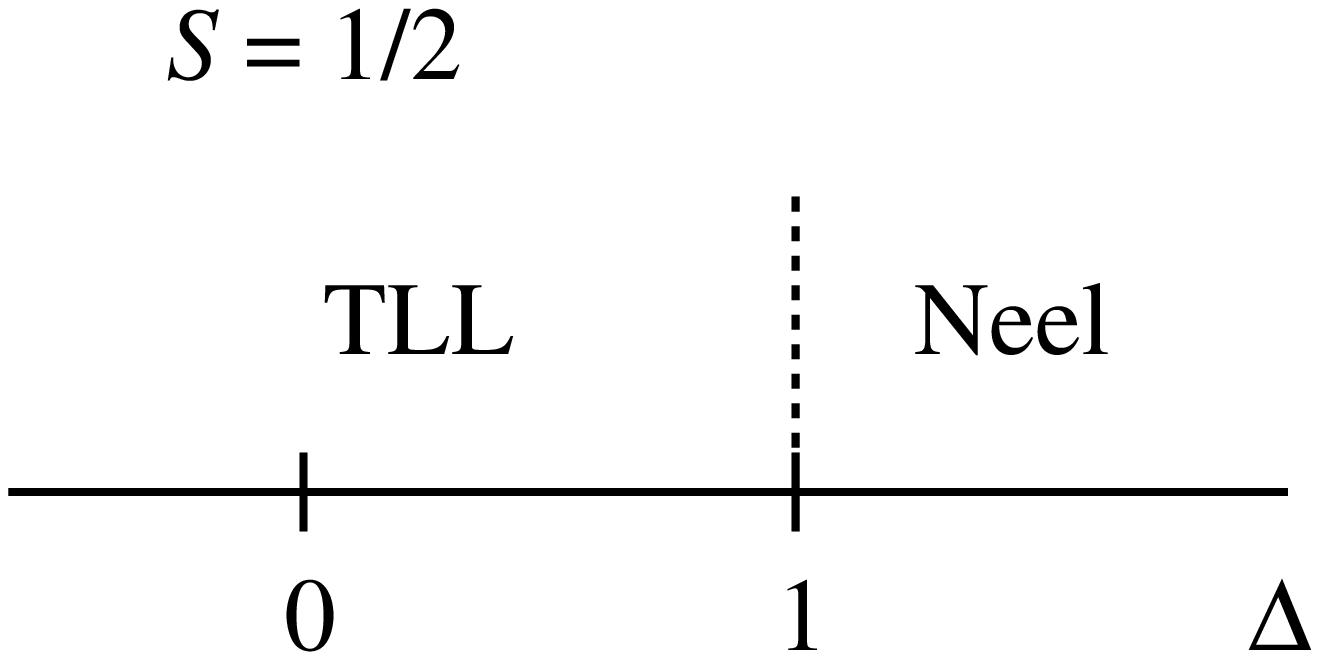}}~~~~~~~~~~
          \scalebox{0.45}{\includegraphics{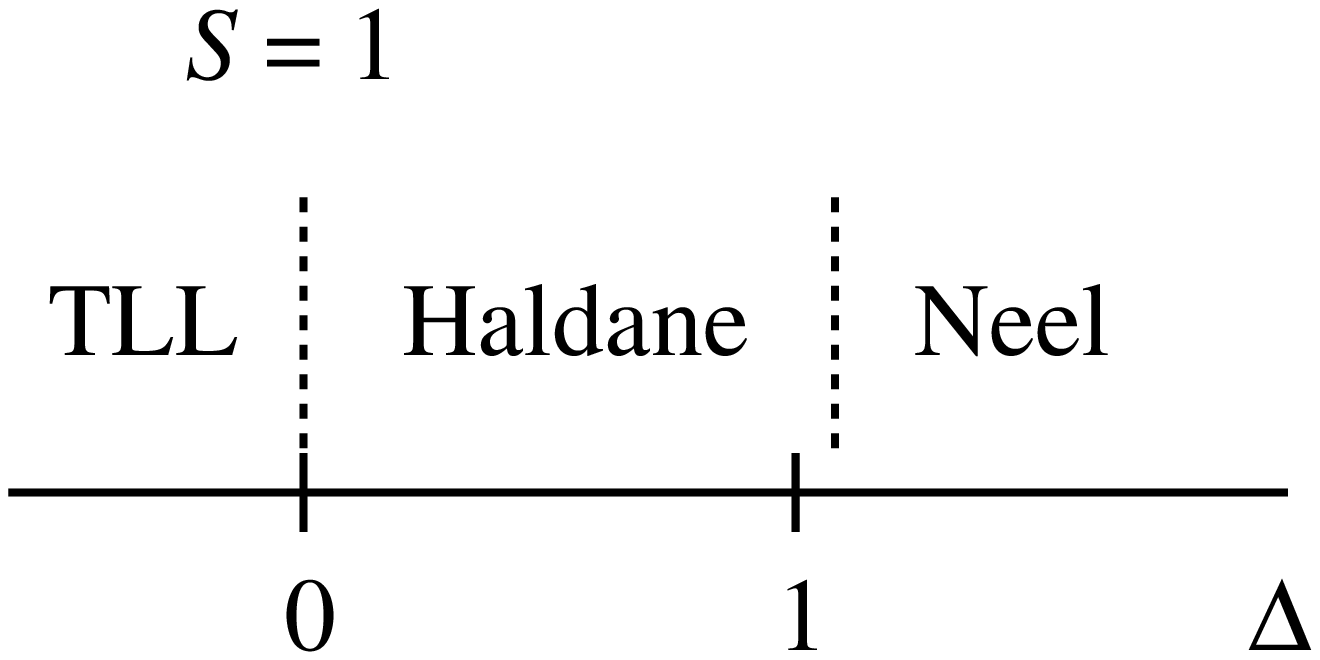}}}
       \caption{Ground state phase diagrams of the $S=1/2$ chain (left panel)
                and the $S=1$ chain (right panel)
                with the $XXZ$ interaction anisotropy.
                We can see that Ising-like anisotropy favors the N\'eel state,
                while $XY$-like anisotropy does the TLL state.}
\end{figure}

Several years go, we have found the {\it inversion phenomena}
between the anisotropies of the interaction and the wave function
in the $S=1/2$ distorted diamond spin chain model \cite{oka-ichi,tokuno-oka}
and $S=1/2$ trimerized spin chain model with the next-nearest-neighbor interactions \cite{oka-ptp}. 
Namely, in some parameter regions,
the TLL state is realized for the Ising-like anisotropy case
and the N\'eel state for the $XY$-like anisotropy case.
These inversion phenomena are considered to be attributed to the interplay
among the geometrical frustration, the trimer nature and the $XXZ$ anisotropy of the
Hamiltonian.
In case of strong geometrical frustrations, 
the $S=1/2$ spin system tends to fall into the dimer state
to avoid the energy loss coming from competing interactions \cite{oka-nomu1}.
But the formation of singlet pairs is incompatible with the trimer nature
of the Hamiltonian.
Thus, for instance, in the $XY$-like anisotropy case,
the spins turn to the $z$-direction to avoid the energy loss, 
because interactions along the $z$-direction is weaker than those in the
$xy$-direction.
Similarly, we can explain the existence of the TLL state
for the Ising-like anisotropy case.

If this physical consideration hits the point,
the inversion phenomena will be observed in many quantum $S=1/2$ spin chain models
having the geometrical frustration, the trimer nature and the $XXZ$ anisotropy.
From this point of view,
we discuss the $S=1/2$ geometrically frustrated 3-leg ladder with the $XXZ$ anisotropy.

\section{Ground-State Phase Diagram of the $S=1/2$ Geometrically Frustrated 3-Leg Ladder}

The $S=1/2$ geometrically frustrated 3-leg ladder with the $XXZ$ anisotropy is expressed by
\begin{eqnarray}
   \cH
   &=& J_1 \sum_{j} \sum_{i=1,2,3} h_{(j,i),(j+1,i)}(\Delta)
      + J_\rr \sum_{j} \sum_{i=1,2} h_{(j,i),(j,i+1)}(\Delta) \nonumber \\
     &&+ J_\rd \sum_{j} \sum_{i=1,2} h_{(j,i),(j+1,i+1)}(\Delta)
        + J_\rd \sum_{j} \sum_{i=2,3} h_{(j,i),(j+1,i-1)}(\Delta)
\end{eqnarray}
where 
\begin{equation}
    h_{(m),(n)} \equiv S_m^x S_n^x + S_m^x S_n^x + \Delta S_m^x S_n^x,~~~\Delta > 0
\end{equation}
Here $S_{j,i}^\mu$ ($\mu = x,y,z$) represents the $\mu$-component of the $S=1/2$ spin operator
at the $j$-th site of the $i$-th chain,
$\Delta$ the $XXZ$ anisotropy parameter.
All the coupling constants, $J_1$, $J_\rr$ and $J_\rd$, are supposed to be positive (antiferromagnetic).
The present model is sketched in Fig.2.
Hereafter we restrict ourselves to the $J_\rr \gg J_1,J_\rd$ case for simplicity,
because the inversion phenomena were so far observed in the strong trimer nature cases.

\begin{figure}[ht]
       \centerline{\scalebox{0.4}{\includegraphics{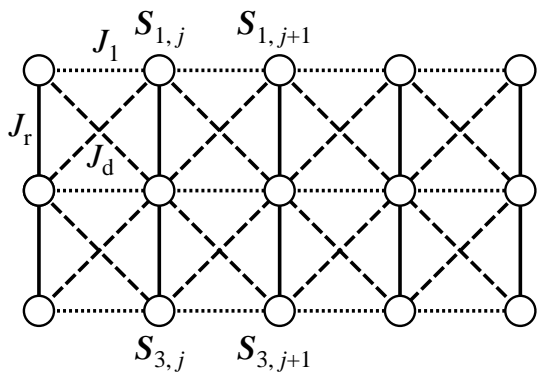}}}
       \caption{Sketch of the present model.
                Horizontal dotted lines denote $J_1$,
                vertical solid lines $J_\rr$, 
                and diagonal broken lines $J_\rd$.
                These three coupling constants are supposed to be positive (antiferromagnetic).
                We can see that there are geometrical frustrations.
                In the absence of $J_\rd$, there is no geometrical frustration.}
\end{figure}

Let us discuss the ground state of our model by use
of the degenerate perturbation theory.
First we consider the 3-spin problem of $\vS_{j,1},\vS_{j,2}$ and $\vS_{j,3}$,
wich can be easily solved.
The ground states of the $j$-th 3-spin cluster are
\newpage
\begin{eqnarray}
    &&\ket{\Up_j}
    \equiv \frac{1}{A}(\ket{\up\up\dn} - a\ket{\up\dn\up} + \ket{\dn\up\up}) 
    \\ \nonumber \\
    &&\ket{\Dn_j}
    \equiv \frac{1}{A}(\ket{\up\up\dn} - a\ket{\up\dn\up} + \ket{\dn\up\up})
\end{eqnarray}
where $\ket{\up\up\dn}$ means $\ket{\up_{j,1}\up_{j,2}\dn_{j,3}}$
for instance, and
\begin{equation}
    A=\sqrt{2+a^2},~~~~~a = \frac{\Delta + \sqrt{\Delta^2 + 8}}{2}
\end{equation}
Typeical values of $A$ and $a$ are as follows:
$A=2,\,a=\sqrt{2}$ for $\Delta =0$,
$A=\sqrt{6},\,a=2$ for $\Delta =1$,
and
$A=\sqrt{6+2\sqrt{3}},\,a=1+\sqrt{3}$ for $\Delta =2$.

Next we consider the effects of $J_1$ and $J_\rd$.
As far as $J_\rr \gg J_1,J_\rd$,
we can take into consideration only these two states, $\ket{\Up_j}$ and $\ket{\Dn_j}$, 
for the $j$-th trimer,
neglecting other 6 states.
For convenience we consider $\ket{\Up_j}$ and $\ket{\Dn_j}$ as 
the up-spin and down-spin states of the pseudo-spin $\vT_j$, respectively.
The interactions between the trimers are expressed as the interactions
between pseudo-spins.
By use of the lowest order perturbation theory with respect to $J_l$ and $J_\rd$,
we obtain
\begin{equation}
    \cHeff
    = \sum_j \left\{
        J_{\rm eff}^\perp \left(T_j^x T_{j+1}^x + T_j^y T_{j+1}^y \right)
        + J_{\rm eff}^z T_j^z T_{j+1}^z \right\}
\end{equation}
where
\begin{equation}
    J_{\rm eff}^\perp
    = \frac{(8a^2+4)J_1 - 16 a J_\rd}{A^4},~~~~~~
    J_{\rm eff}^z
    = \frac{(3a^4-4a^2+4)J_1 + 4a^2(a^2-2)J_\rd}{A^4}
    \label{eq:Jeff}
\end{equation}

When $\Delta=1$, we see $J_{\rm eff}^\perp = J_{\rm eff}^z = J_1-(8/9)J_\rd$.
Then the ground state of $H_{\rm eff}$ is either the ferromagnetic state
or the TLL state depending on whether $J_\rd > (9/8)J_1$ or $J_\rd < (9/8)J_1$.
The ferromagnetic state of the $\vT$-system corresponds to
the ferrimagnetic state of the $\vS$-system,
because all the 3-spin clusters are in the same state, $\ket\Up$ or $\ket\Dn$.
This ferrimagnetic state has the magnetization of $M_{\rm s}/3$,
where $M_{\rm s}$ is the saturation magnetization.
For the general case ($\Delta \ne 1$),
the ground states of $H_{\rm eff}$ are known from $J_{\rm eff}^\perp$ and $J_{\rm eff}^z$,
and are translated into the $\vS$-system language as 
\begin{equation}
    \begin{matrix}
                                                        &\vT{\rm -picture}  &\vS{\rm -picture}  \\
        J_{\rm eff}^z > |J_{\rm eff}^\perp|      \hfill &\hbox{\rm N\'eel}  &\hbox{\rm N\'eel}  \\
        |J_{\rm eff}^z| < |J_{\rm eff}^\perp|     \hfill &{\rm TLL}          &{\rm TLL}          \\
        J_{\rm eff}^z < - |J_{\rm eff}^\perp|  \hfill &{\rm ferro}        &{\rm ferri}~(M_\rS/3)        \hfill \\
    \end{matrix}    
    \label{eq:condition}
\end{equation}

The phase diagrams for three cases, $\Delta=1,\ 0.5\mbox{~and~}2.5$,
near the truncation point $\tilde J_1 = \tilde J_\rd =0$ are shown
in Fig. 3,
where $\tilde J_1 \equiv J_1/J_\rr,\ \tilde J_\rd \equiv J_\rd/J_\rr$.
The remarkable nature of the $\Delta = 0.5$ case (center panel of Fig. 3)
is the existence of the
N\'eel region,
although the interaction anisotropy is $XY$-like. 
Similarly, the TLL state is realized even for the $\Delta > 1$ case (right panel of Fig. 3)
in spite of the Ising-like anisotropy.
These are the {\it inversion phases}.

\begin{figure}[ht]
   \begin{center}
      \scalebox{0.25}{\includegraphics{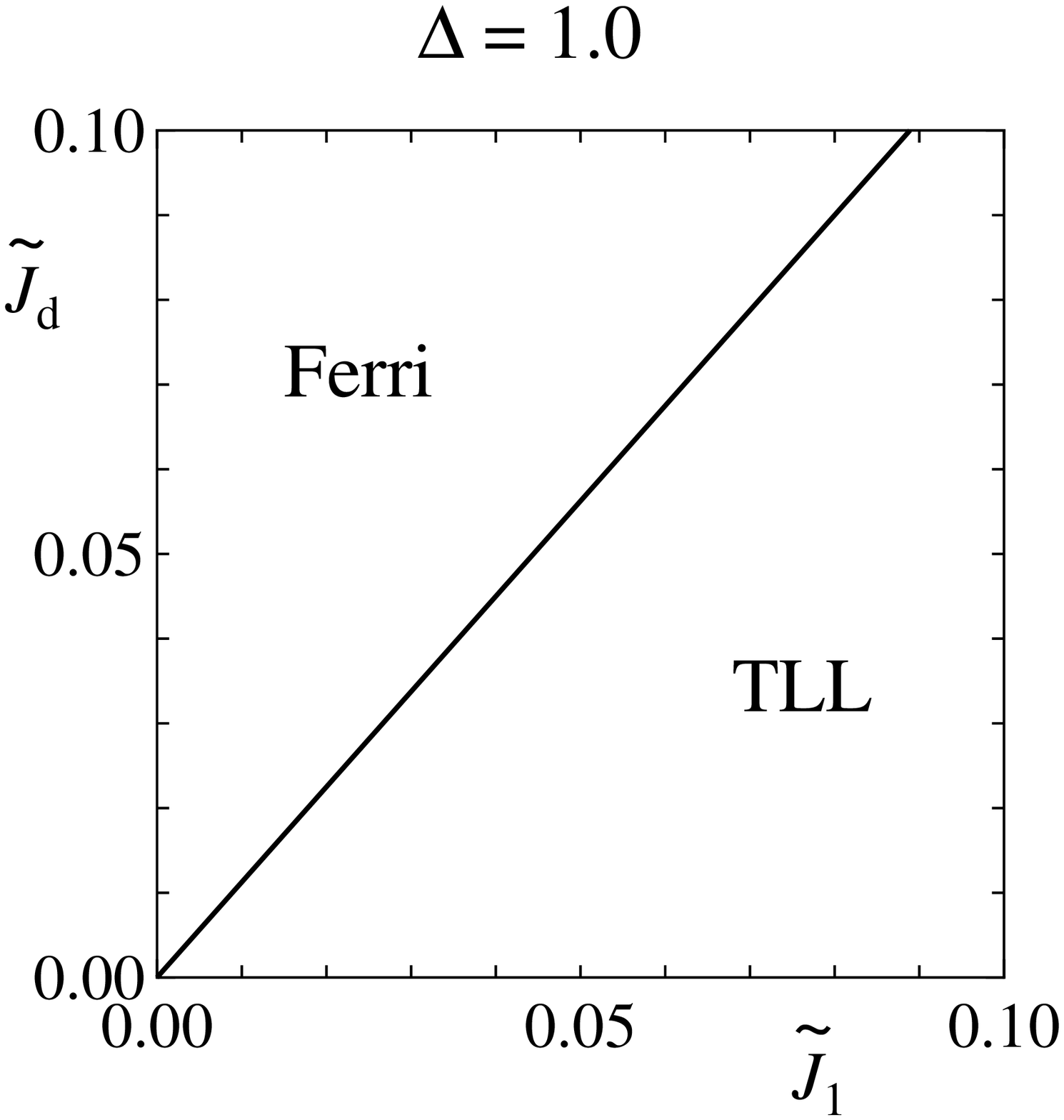}}~~~
      \scalebox{0.25}{\includegraphics{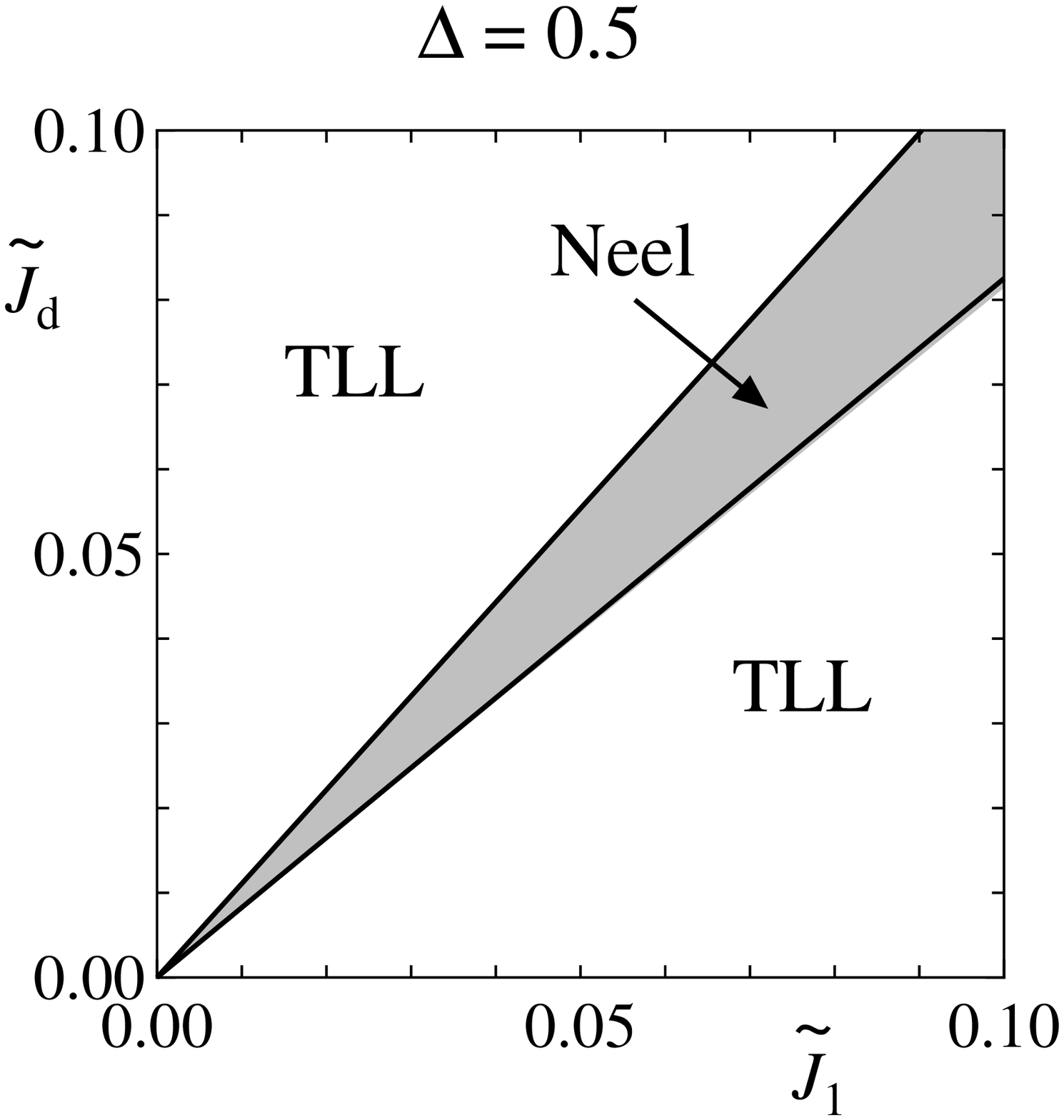}}~~~
      \scalebox{0.25}{\includegraphics{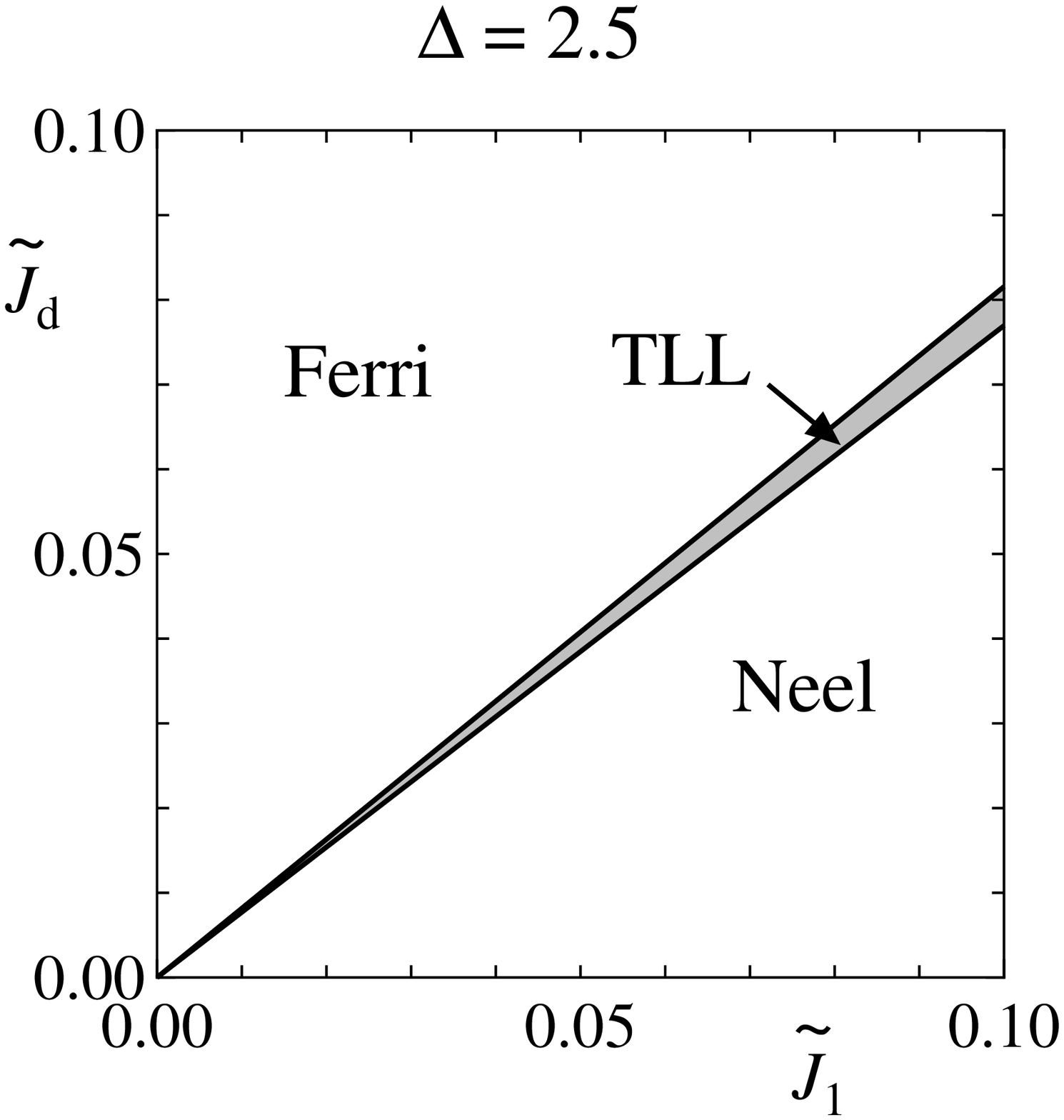}}
   \end{center}
   \caption{Ground-state phase diagrams on the $\tilde J_1 - \tilde J_\rd$ plane for
            the $\Delta = 1.0$, $\Delta=0.5$ and $\Delta = 2.5$ cases,
            where $\tilde J_1 \equiv J_1/J_\rr,\ \tilde J_\rd \equiv J_\rd/J_\rr$.
            The shaded areas are the inversion regions.}
\end{figure}

We have also performed the numerical diagonalization by use of the
Lancz\"os method and analyzed the numerical data by the
level spectroscopy method \cite{oka-nomu1,oka-nomu2,oka-nomu3}.
For instance, when $\Delta = 0.5$ (center panel case of Fig.3),
from the numerical data,
the boundary lines between the TLL and N\'eel phase are
$\tilde J_\rd \simeq \tilde 0.8252 J_1$ and $\tilde J_\rd \simeq \tilde 1.106 J_1$.
These results very well agrees with the perturbation results
$\tilde J_\rd = \tilde 0.8252 J_1$ and $\tilde J_\rd = \tilde 1.107 J_1$
obtained from eqs.({\ref{eq:Jeff}}) and (\ref{eq:condition}).

\section{Discussoin and Concluding Remarks}

We have shown that there exist inversion regions in the
$S=1/2$ geometrically frustrated 3-leg ladder with the $XXZ$ interaction anisotropy
by use of the degenerate perturbation theory,
as expected from the physical consideration described in \S1.
We have also checked our analytical conclusion by numerical methods.
The key to the inversion phenomena is the form of $J_{\rm eff}^\perp$
and $J_{\rm eff}^z$.
If $J_\rd =0$,
it is easy to see that
$J_{\rm eff}^\perp > J_{\rm eff}^z > 0$ for $\Delta < 1$
and $0< J_{\rm eff}^\perp < J_{\rm eff}^z$ for $\Delta > 1$.
Thus the inversion phenomena do not occur in case of $J_\rd =0$.
This fact implies that the geometrical frustration is
essential to the realization of the anisotropy inversion
phenomena.

The contents of this paper strongly suggest
that the inversion phenomena are popular to
the $=1/2$ $XXZ$ chain and ladder models with the geometrical frustration and the trimer nature.

\section*{Acknowledgments}

This work was partly supported by Grants-in-Aid (Nos. 23340109 and 23540388)
for Scientific Research from the Ministry of Education, Culture, Sports, Science and Technology of Japan.

\end{document}